# Optimising Blockchain Scalability for Real-Time IoT Applications

Hasan Mahmud Rhidoy [a], Mahdi H. Miraz [a,b,c,*], Iftekhar Salam [a,d]

[a] *School of Computing and Data Science, Xiamen University Malaysia, Malaysia*
[b] *School of Computing, Wrexham University, United Kingdom*
[c] *Faculty of Computing, Engineering and Science, University of South Wales, United Kingdom*
[d] *Faculty of Science and Information Technology, Daffodil International University, Bangladesh*

Correspondence: *m.miraz@ieee.org*

*Abstract*—The convergence of blockchain and the Internet of Things (IoT) enables secure, decentralised, and verifiable data exchange across distributed smart environments. However, traditional blockchain frameworks suffer from inherent scalability constraints, limited throughput, and high latency, which conflict with the stringent real-time requirements of IoT applications such as industrial automation, intelligent healthcare, and smart transportation. These systems demand ultra-low latency, high transaction throughput, lightweight computation, and efficient resource utilisation. This review provides a comprehensive, structured analysis of state-of-the-art scalability solutions specifically adapted to blockchain-enabled IoT. The discussion encompasses Layer 1 enhancements, Layer 2 off-chain processing, sharding-based parallelisation, integration of edge and fog computing, and hybrid consensus mechanisms. For each approach, the review highlights operational principles, performance benefits, trade-offs in decentralisation and security, and suitability for latency-sensitive deployments. Furthermore, real-time quality-of-service considerations are examined to understand how scalability strategies impact system responsiveness, energy efficiency, and data integrity. Key open challenges, including the scalability-security trade-off, privacy preservation, interoperability, and sustainable resource management, have been identified as persistent barriers to large-scale adoption. Finally, the review outlines future research directions, emphasising adaptive and AI-driven consensus algorithms, quantum-safe cryptographic models, the convergence of blockchain with 5G/6G networks, and edge intelligence. By consolidating diverse technical insights and emerging trends, this work serves as a timely reference for developing scalable, secure, and sustainable blockchain architectures for real-time IoT applications.

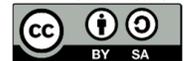

## I. INTRODUCTION

Internet of Things (IoT) is expanding at an exponential rate, connecting billions of devices in multifaceted applications, including smart healthcare [1], industrial automation, intelligent transport systems, and many more [2], [3]. These IoT applications often generate massive real-time data streams that must be processed with very low latency and high efficiency to enable rapid decision-making and improved system performance for both applications and users [4], [5]. Despite their potential, IoT systems frequently face challenges due to limited resources and the absence of unified standards, which lead to vulnerabilities in security, data integrity, and trust. Integrating blockchain technology offers promising solutions to these issues by providing decentralised, immutable, and tamper-proof ledgers, with smart contracts enabling autonomous data exchange [6]. This combination enhances privacy and mitigates single points of failure that typically arise in centralised IoT frameworks [7].

However, the traditional blockchain architecture has inherent limitations, including scalability and latency. The use of consensus mechanisms, such as Proof of Work (PoW) and Proof of Stake (PoS), complicates matters by introducing latency and computational costs, which are unsuitable for the latency-sensitive, real-time nature of IoT applications [8], [9]. Moreover, most IoT devices have limited computational power and storage capacity, which poses a major challenge for deploying a full blockchain [10]. Recent studies have proposed various scalability approaches specific to the integration of blockchain and IoT, including Layer-2 solutions, sharding, lightweight consensus protocols, and a hybrid blockchain–edge-computing paradigm, in which computation is moved closer to devices to mitigate latency [11]-[14]. Nonetheless, although these developments have already occurred, the authors are not aware of any comprehensive survey that directly addresses the optimisation of blockchain scalability for real-time IoT applications.

This review addresses this gap by analysing state-of-the-art blockchain scalability solutions with a specific focus on real-time IoT constraints, including ultra-low latency, Quality of Service (QoS), and deterministic responsiveness. Unlike existing surveys, this study evaluates how scalability techniques affect latency-sensitive and time-critical IoT applications, positioning scalability optimisation as a key enabler of real-time and QoS-aware IoT systems.

## II. MATERIALS AND METHODS

### A. Literature Search Strategy and Selection Criteria

This review adopts a systematic and structured literature search strategy to identify, analyse, and synthesise state-of-the-art research on blockchain scalability optimisation for real-time IoT applications. The primary objective of the search process was to identify peer-reviewed studies that explicitly address scalability, latency, throughput, and resource-efficiency challenges arising from the integration of blockchain with IoT systems.

Relevant literature was collected from well-established scientific databases, including IEEE Xplore, Scopus, Web of Science, ScienceDirect, and Google Scholar, which collectively cover a broad spectrum of high-impact journals and conference proceedings in blockchain, distributed systems, and IoT research. The search was conducted using combinations of keywords such as "blockchain scalability", "blockchain IoT integration", "real-time IoT", "low-latency blockchain", "edge computing blockchain", "sharding", "Layer 2 solutions", "consensus mechanisms", and "QoS-aware IoT systems". Boolean operators and keyword variations were applied to ensure comprehensive coverage of relevant studies.

To ensure the review's relevance and timeliness, the search primarily focused on publications from 2018 to 2025, reflecting the period during which blockchain–IoT convergence and scalability optimisation gained significant research momentum. Earlier foundational works were selectively included where necessary to establish background concepts related to blockchain architecture and consensus mechanisms. A set of inclusion and exclusion criteria was applied to filter the retrieved studies. Included papers were required to:

- Explicitly address blockchain scalability, performance optimisation, or architectural enhancements.
- Consider IoT or cyber–physical systems as a target application domain, and discuss performance metrics relevant to real-time systems, such as latency, throughput, energy efficiency, or Quality of Service (QoS).

Studies were excluded if they focused solely on cryptocurrency trading, lacked technical depth, or did not consider scalability or real-time constraints. Non-peer-reviewed articles, tutorial-only papers, and non-English publications were also excluded to maintain academic rigor.

Following the initial screening based on titles and abstracts, a full-text assessment was conducted to ensure alignment with the review objectives. The final set of selected studies was then analysed and categorised according to their proposed scalability approaches, including Layer 1 enhancements, Layer 2 off-chain mechanisms, sharding techniques, integration of edge and fog computing, and hybrid consensus models. This systematic selection process ensures that the review provides a comprehensive, balanced, and up-to-date perspective on scalability optimisation strategies for blockchain-enabled real-time IoT systems.

### B. Literature Search Strategy and Selection Criteria

The reviewed studies were categorised into Layer 1 enhancements, Layer 2 off-chain solutions, sharding, edge/fog integration, and hybrid consensus mechanisms based on their scalability approach. Each category was comparatively analysed using real-time IoT performance criteria, including latency, throughput, energy efficiency, and decentralisation trade-offs, as summarised in Table I. This framework enables a systematic comparison of optimisation strategies under real-time constraints. It also supports the identification of open challenges and research gaps discussed in subsequent sections.

The remainder of this paper is organised as follows. Section III presents the fundamental concepts of blockchain and IoT, outlining their core architectures and roles in supporting secure and efficient system operation. Section IV reviews and analyses blockchain scalability techniques, including Layer 1 and Layer 2 solutions, sharding mechanisms, and edge computing integration, with emphasis on their applicability to IoT environments. Section V discusses real-time considerations, focusing on latency, throughput, and Quality of Service (QoS) requirements for time-critical IoT applications.

### C. Fundamentals of Blockchain and IoT

Blockchain is a distributed, decentralised, ledger-based technology that records transactions securely, irrevocably, and auditably by combining cryptographic protocols and consensus mechanisms. It enables peer-to-peer interaction among untrusted parties by eliminating the need for a central authority, establishing trust through data structures, cryptographic algorithms, and consensus mechanisms. Blockchain architecture consists of two interdependent components: data structures, which ensure the integrity and durability of data, and consensus algorithms, which facilitate agreement among distributed nodes on the current state of the ledger [15]. From a data perspective, the blockchain is a chain of blocks, each containing a list of verified transactions, a timestamp, and a cryptographic hash that links it to the previous block [16]. The hash-chaining attribute guarantees that tampering with any block will invalidate all other blocks and require recomputing the cryptographic hash of each block, making such tampering computationally infeasible. The transactions in each block are organised in a Merkle tree, an efficient binary hash tree that enables verification and integrity checks without duplicating the data [17].

Beyond its structural components, blockchain can be understood as a trust-enabling coordination layer for distributed systems. Instead of relying on a central intermediary to validate and store information, blockchain enables independent participants to collectively maintain a shared, consistent system state. This shift is particularly valuable in environments where stakeholders do not fully trust each other, yet must exchange data or value reliably. By embedding verification, ordering, and accountability directly

312

into the protocol, blockchain transforms trust from an organizational assumption into a technical property. In practice, this means that system behavior becomes transparent, traceable, and resilient to unilateral manipulation.

Such characteristics are increasingly important in data-driven ecosystems, where accountability and provenance are as critical as performance, especially when systems scale across organizational and geographical boundaries [15], [16].

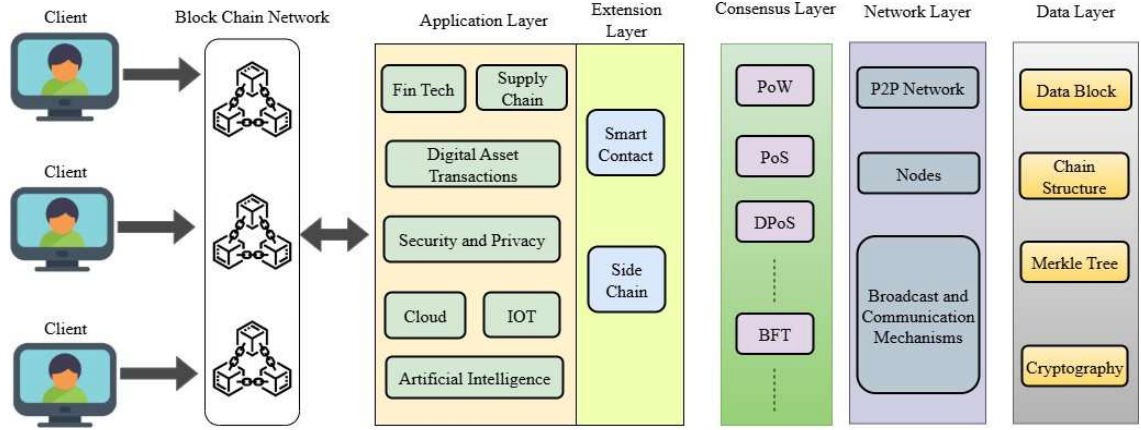

Fig. 1 Architecture of Blockchain Technology

In this setup, lightweight clients, e.g., resource-constrained IoT devices, do not store the entire blockchain but can simply verify the existence of a transaction. Data important to the integrity of the block are incorporated into the block header: the hash of the previous block, the Merkle root, a timestamp, and other parameters specific to the chosen consensus algorithm. By adopting such a combination, the structural features provide tamper evidence, auditability, and distributed verification. Consensus algorithms are responsible for ensuring that all participants in the network agree on the contents and ordering of transactions, even in the presence of faults or malicious behaviour [18]. Proof-of-work (PoW) is the earliest and most widely recognised consensus protocol, requiring participants to solve computational puzzles before appending a block to the chain. While offering robust resistance to Sybil attacks [19], PoW is energy-intensive and exhibits low throughput, making it less suitable for latency-sensitive IoT scenarios. Proof-of-Stake (PoS) addresses some of these limitations by selecting validators based on their economic stake in the network, thereby improving efficiency and reducing energy consumption [20], [21].

Practical Byzantine Fault Tolerance (PBFT) employs a leader-based voting approach to achieve consensus with low latency in permissioned environments, a property beneficial for industrial IoT systems but with scalability limitations due to communication overhead. More recent approaches, such as Delegated Proof of Stake (DPoS) and Proof of Authority (PoA), enhance scalability by reducing the number of nodes participating in block validation, whilst Directed Acyclic Graph (DAG)-based protocols, such as IOTA's Tangle, allow for parallel transaction processing, improving throughput for high-frequency IoT data streams [22]. Figure 1 illustrates the fundamental architecture of blockchain technology. The combination of immutable, verifiable data structures and consensus fault tolerance that blockchain offers enables trust, transparency, and resilience in decentralised systems.

*D. IoT Architecture and Real-Time Requirements*

The Internet of Things (IoT) is typically structured as a layered architectures that coordinate interactions between physical devices and digital services. The perception layer comprises sensors and actuators that capture environmental data but are constrained in energy, computation, and storage [23]. The network layer forwards this data using heterogeneous technologies such as Wi-Fi, 5G, Zigbee, and LoRaWAN, yet often encounters congestion and reliability issues. Above this, the middleware (or business) layer handles data processing and service orchestration, whilst the application layer delivers domain-specific functions across healthcare, industry, and transportation [24]-[26].

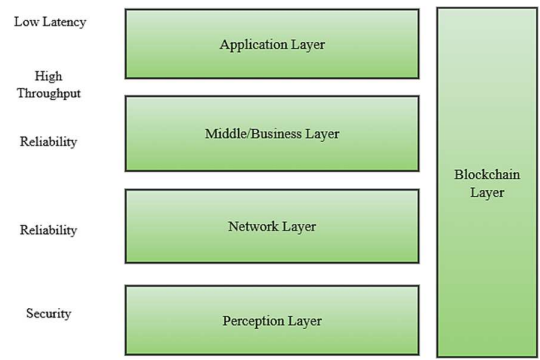

Fig. 2 IoT Architecture and Real-Time Requirements with Blockchain

A defining requirement of IoT systems is real-time performance: applications such as autonomous driving, smart grids, and telemedicine demand ultra-low latency, high throughput, reliability, scalability, and robust security. Even minor delays or communication faults can jeopardise safety or efficiency [27]. Whilst blockchain brings transparency, immutability, and decentralised trust, its consensus overhead and storage requirements often clash with these strict constraints. Bridging blockchain scalability with real-time IoT thus remains a central research challenge. The interplay amongst IoT layers, blockchain integration, and real-time requirements is depicted in Figure 2.

*E. Scalability Challenges of Blockchain in IoT*

Scalability is one of the most critical obstacles when applying blockchain to IoT systems. Traditional blockchain platforms, such as Bitcoin and Ethereum, were not designed to handle the massive transaction volume generated by billions of IoT devices. Their limited throughput, typically a few transactions per second, contrasts sharply with the high-frequency data streams in real-time IoT environments, creating bottlenecks that hinder timely processing and decision-making [28].

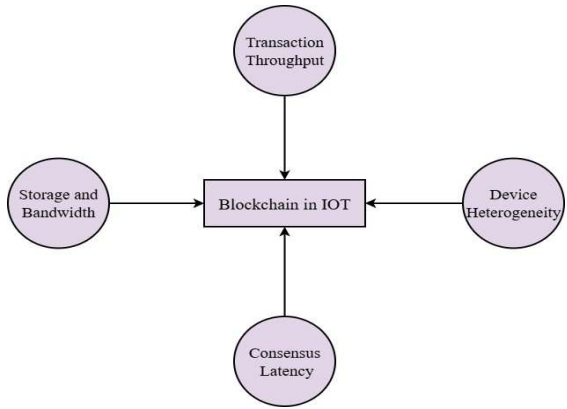

Fig. 3 Scalability Challenges of Blockchain in IoT

Another key challenge arises from consensus mechanisms. Protocols such as Proof-of-Work (PoW) or Proof-of-Stake (PoS) introduce delays in transaction confirmation that are incompatible with the millisecond-level responsiveness required in many IoT applications. The computational intensity and energy consumption of these mechanisms also exceed the capabilities of resource-constrained IoT devices [29]. Storage and bandwidth requirements further complicate the integration. As each blockchain node is expected to maintain a full copy of the ledger, the continuous data generation from IoT devices can quickly overwhelm both storage capacity and communication networks. This becomes particularly problematic in large-scale deployments where maintaining a synchronised distributed ledger demands substantial resources [30].

Finally, heterogeneity and intermittency in IoT environments exacerbate scalability issues. Devices vary in processing power, connectivity, and energy supply, while many operate intermittently to conserve resources. Ensuring secure, real-time blockchain participation across such diverse nodes is inherently difficult [31]. As illustrated in Figure 3, these scalability challenges can be grouped into four main categories: transaction throughput, consensus latency, storage and bandwidth overhead as well as device heterogeneity. Overcoming these barriers is essential before blockchain can be widely adopted for real-time IoT applications, motivating the range of optimisation techniques reviewed in the next section.

*F. Scalability Techniques for Blockchain in IoT*

A variety of scalability techniques have been proposed to address the mismatch between blockchain performance and the stringent requirements of real-time IoT systems. These solutions can broadly be grouped into Layer 1 improvements, Layer 2 solutions, sharding approaches, edge and fog computing integration, and hybrid consensus mechanisms as illustrated in Figure 4.

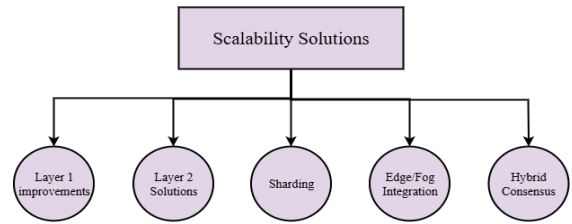

Fig. 4 Taxonomy of Scalability Techniques for Blockchain in IoT

*G. Layer 1 Improvements*

However, these approaches often involve trade-offs between scalability, decentralisation, and security. For instance, increasing block size can improve throughput but also lengthens propagation time, which risks creating forks and weakening consensus security. Shortening block intervals similarly increases the likelihood of orphaned blocks and undermines stability. Permissioned consensus algorithms, such as PBFT and PoA, achieve low latency by limiting participation to trusted nodes, thereby reducing decentralisation and openness. DAG-based protocols (e.g., IOTA's Tangle) enhance parallelism, but their lighter consensus assumptions may make them more vulnerable to certain attacks [32].

*H. Layer 2 Solutions*

Layer 2 solutions reduce the computational and storage burden on the main blockchain by processing the transactions off-chain. Examples include payment channels such as the Lightning Network [33], state channels and sidechains. In these systems, frequent microtransactions are processed off-chain, and only the final state is recorded on-chain, thereby reducing latency and significantly improving throughput. As illustrated in Figure 5, Layer 2 mechanisms enable off-chain microtransactions whilst ensuring that final settlements are securely anchored to the main blockchain.

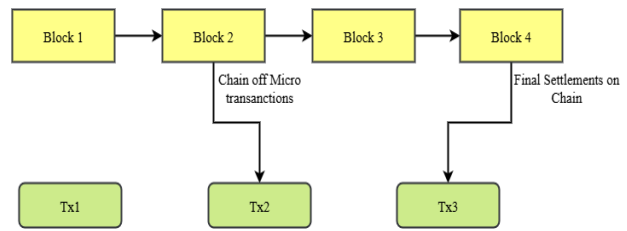

Fig. 5 Layer 2 Solutions using Off-chain Channels and Sidechains in IoT

This design is particularly suitable for IoT scenarios involving continuous data exchanges or device-to-device micropayments, such as energy trading in smart grids and dynamic resource sharing in industrial IoT. The main advantage of Layer 2 solutions lies in their ability to provide near-instant responses and high scalability without altering the underlying blockchain protocol. However, they also introduce additional complexity in dispute resolution, as off-chain activities must be securely synchronised with the main chain. Furthermore, reliance on off-chain networks may expose vulnerabilities if those auxiliary systems are compromised [34].

*I. Sharding*

Sharding divides the blockchain into smaller groups of nodes, known as shards, each of which processes its own set of transactions in parallel. This partitioning significantly increases throughput because multiple shards can validate and store transactions concurrently. For IoT environments, sharding is more favourable because large-scale device networks can be organised by function, region or application domain, thereby distributing the computational load more evenly. The primary advantage of sharding lies in its potential to achieve near-linear scalability as the number of shards increases. However, it also introduces new vulnerabilities, particularly in cross-shard communication, which is required for transactions that span multiple shards. If not carefully designed, this process can create synchronisation delays and open avenues for targeted attacks on individual shards [35].

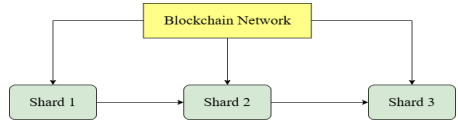

Fig. 6 Sharding in Blockchain Networks for IoT Scalability

As shown in Figure 6, sharding allows the blockchain network to process transactions across multiple shards in parallel, while cross-shard coordination ensures consistency. This makes sharding a promising, though technically complex, solution for addressing the scalability needs of real-time IoT systems.

*J. Edge and Fog Computing Integration*

Integrating blockchain with edge and fog computing places computation and validation closer to IoT devices, thereby reducing reliance on remote cloud or centralised infrastructure. Edge and fog nodes can execute lightweight consensus, store partial ledgers, and validate local transactions before synchronising with the core blockchain. This design minimises communication delay and improves responsiveness, making it highly suited for time-critical IoT scenarios such as autonomous vehicles, industrial automation, and healthcare monitoring. The major advantage of this approach is its ability to reduce latency and network congestion by offloading tasks from the central blockchain. Moreover, local processing supports scalability by distributing workload across multiple nodes near the data source. However, edge and fog computing resources remain constrained relative to cloud systems, and the security of distributed edge nodes remains a persistent concern [36].

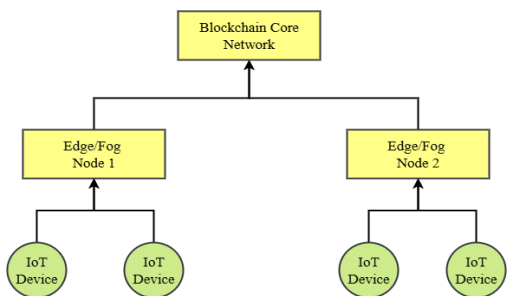

Fig. 7 Edge and Fog Computing Integration with Blockchain for IoT

As illustrated in Figure 7, IoT devices interact with nearby edge and fog nodes, which perform preliminary validation and forward only essential data to the blockchain core. This layered approach effectively reduces latency and enhances real-time performance whilst maintaining blockchain's security and transparency.

*K. Hybrid Consensus Mechanisms*

Hybrid consensus mechanisms combine the strengths of multiple protocols to balance scalability, security and efficiency in blockchain systems. For example, IoT-oriented networks may employ Delegated Proof of Stake (DPoS) to achieve fast transaction validation, whilst using Byzantine Fault Tolerance (BFT) to ensure strong finality. Similarly, Proof of Authority (PoA) can be integrated into permissioned settings to further reduce latency and energy consumption. The advantage of hybrid consensus lies in its flexibility: it can be tailored to the heterogeneous requirements of IoT ecosystems, where some applications demand ultra-low latency while others prioritise fault tolerance or decentralisation. However, this approach increases architectural complexity and may create governance challenges in coordinating multiple protocols [37].

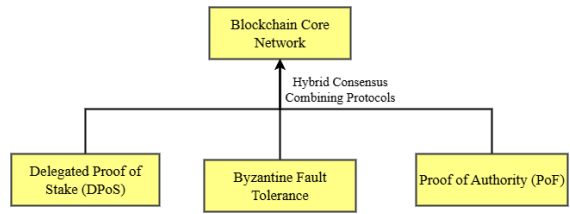

Fig. 8 Hybrid Consensus Mechanisms for Scalable Blockchain in IoT

As depicted in Figure 8, different consensus methods can be combined into a hybrid model that supports the blockchain network, providing scalability without fully sacrificing decentralisation or security. This makes hybrid consensus a practical and adaptable option for real-time IoT systems.

*L. Real-Time Considerations*

Real-time performance is a defining requirement of many IoT applications, where even millisecond delays can compromise safety or efficiency. Blockchain integration introduces additional latency through consensus mechanisms and block confirmation times, which often conflict with the strict responsiveness required in domains, such as healthcare, smart grids and autonomous vehicles. Throughput limitations further hinder the ability to support massive device interactions, whilst network conditions, such as jitter and packet loss, can aggravate delays. A key trade-off exists amongst decentralisation, security, and speed: highly secure consensus protocols tend to be slower, whereas lightweight mechanisms improve latency at the expense of robustness. Quality of Service (QoS) metrics—including latency, reliability and fault tolerance—must, therefore, be carefully balanced when adapting blockchain to IoT [38, 39].

As illustrated in Figure 9, the integration of blockchain into real-time IoT requires balancing decentralisation, security, and speed. IoT systems naturally demand low latency and high throughput, but without compromising trust and resilience. Achieving this balance necessitates protocol optimisations and architectural enhancements, as explored in the preceding sections.

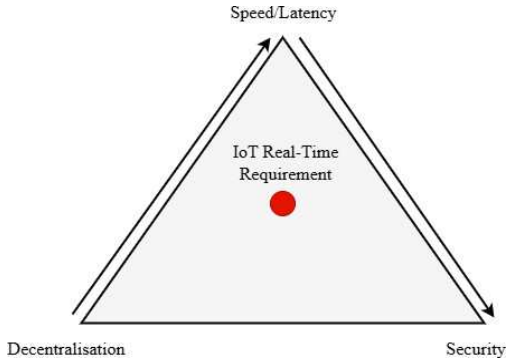

Fig. 9 Blockchain Trilemma in Real-time IoT Systems: Latency, Security, and Decentralisation.

## III. RESULTS AND DISCUSSION

Despite significant progress in developing scalability solutions, several challenges remain before blockchain can be seamlessly integrated into real-time IoT environments. One critical issue is the trade-off between scalability and security. Many lightweight consensus mechanisms or off-chain solutions improve latency but may reduce fault tolerance or introduce vulnerabilities. Ensuring that optimisations do not compromise security remains an open research direction.

Another major challenge is interoperability across diverse blockchain platforms and IoT ecosystems. Devices and networks often rely on heterogeneous communication protocols, making standardisation essential for large-scale deployment [40]. One potential approach to improving interoperability is the use of atomic swaps, which allow direct cross-chain transactions without intermediaries. Atomic swaps ensure trustless exchange of assets across heterogeneous blockchain platforms, reducing reliance on centralised exchanges and improving flexibility in IoT ecosystems [41]-[44]

Data management and privacy are also pressing concerns. While blockchain ensures transparency and immutability, storing sensitive IoT data directly on-chain may violate privacy requirements and overwhelm limited storage resources [45]. Moreover, privacy regulations such as the EU's General Data Protection Regulation (GDPR) conflict with blockchain's immutability. This creates tension between regulatory compliance and the permanent on-chain storage of IoT data, necessitating privacy-preserving solutions, such as off-chain storage, encryption, and selective disclosure mechanisms.

Energy efficiency and sustainability also present major obstacles. Many IoT devices are energy-constrained, and consensus protocols often impose heavy computational loads. Designing energy-aware blockchain frameworks tailored to resource-constrained IoT nodes is therefore essential to ensure sustainable deployment in large-scale, real-time environments.

Domain-specific implications: These challenges manifest differently across various IoT domains. In *healthcare*, real-time monitoring and telemedicine intensify the latency–security trade-off: confirmation delays must be minimised whilst preserving data integrity. Strict privacy and compliance requirements limit on-chain storage, motivating selective disclosure and off-chain data handling to remain aligned with regulatory frameworks [45]. Interoperability across heterogeneous medical devices and systems also demands common interfaces and standards for safe, large-scale deployment [40].

In Industrial IoT (IIoT), production cells and safety-critical control loops require predictable latency and low jitter. Sharding or hierarchical partitioning can raise throughput but introduces cross-partition coordination that must be bounded to maintain deterministic operation [40]. Consortium-style governance (permissioned validators) can balance decentralisation with auditability, whilst energy-aware consensus reduces the operational cost of continuously active nodes.

In smart transport, highly mobile V2X scenarios combine bursty, city-scale workloads with frequent handovers. Off-chain exchanges can reduce on-chain latency for micropayments and access control, whilst periodic settlement preserves integrity. Interoperability across subsystems, such as traffic management, charging, and tolling, benefits from cross-chain mechanisms to avoid central brokers and enable trustless value exchange [41]–[44]. Privacy protection and minimal on-chain personal data are equally important for compliance and scalability [45].

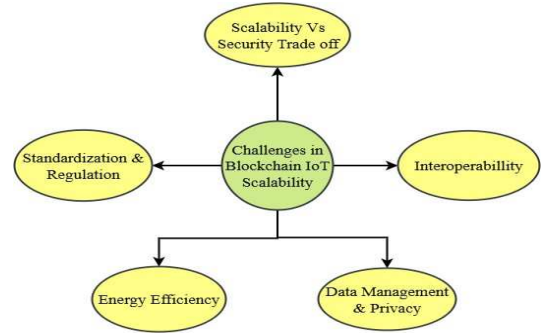

Fig. 10 Challenges and Open Issues in Blockchain for Real-Time IoT

Overall, these domain-specific implications highlight that achieving scalable and secure blockchain integration in real-time IoT requires context-aware optimisation strategies that balance latency, throughput, energy efficiency, and decentralisation within each application environment.

TABLE I
COMPARATIVE EVALUATION OF BLOCKCHAIN SCALABILITY TECHNIQUES FOR REAL-TIME IOT

| Scalability Technique | Latency Reduction | Throughput Improvement | Energy Efficiency | Decentralisation Trade-Off |
|---|---|---|---|---|
| Layer 1 Enhancements (PoS, PBFT, DAG) [15], [21], [22] | Moderate | Moderate | Moderate | Moderate (limited validator set) |
| Layer 2 Off-Chain Channels [9]-[33] | High | High | High | Medium (off-chain trust dependency) |
| Sharding-Based Parallelisation [31], [35] | Moderate | Very High | Moderate | Medium (cross-shard coordination risk) |
| Edge / Fog Computing Integration [12], [13], [36] | Very High | Moderate | High | Medium (permissioned edge nodes) |
| Hybrid Consensus Mechanisms (DPoS + BFT / PoA) [15], [21], [37] | High | High | Moderate–High | Medium (governance complexity) |

## IV. CONCLUSIONS

This review has examined the scalability challenges of blockchain in real-time IoT environments and analysed the main optimisation techniques proposed to address them. Layer 1 improvements, Layer 2 solutions, sharding, edge and fog integration, and hybrid consensus mechanisms each provide valuable strategies for enhancing throughput and reducing latency. However, none of these approaches fully resolves the tension between scalability, decentralisation, and the strict real-time requirements of IoT applications. As summarised in Table 1, these scalability techniques exhibit distinct trade-offs among latency reduction, throughput, energy efficiency, and decentralisation. As highlighted in this study, the integration of blockchain into IoT must account not only for performance but also for device heterogeneity, resource limitations, and energy efficiency.

Looking forward, several promising research directions emerge. First, lightweight and adaptive consensus protocols are needed to reduce computational overhead whilst maintaining security guarantees. Integration with next-generation networks, such as 5G and 6G, combined with edge intelligence, can further reduce latency and improve scalability for large-scale deployments. Another critical direction is the development of interoperable frameworks and standards to ensure seamless interaction between heterogeneous IoT devices and blockchain platforms. Finally, privacy-preserving and energy-aware blockchain architectures must be designed to support sustainable and secure IoT ecosystems. In conclusion, achieving a scalable blockchain for real-time IoT will require hybrid and adaptive approaches that balance performance, trust, and efficiency. Collaborative efforts across research, industry, and standardisation bodies are essential to transform these conceptual solutions into practical, globally deployable systems that support the next generation of IoT applications.

## ACKNOWLEDGEMENT

This research is financially supported by Xiamen University Malaysia (Project Codes: XMUMRF/2021-C8/IECE/0025 and XMUMRF/2022-C10/IECE/0043).